\documentclass{sigplanconf}

\usepackage{xspace}
\usepackage{amssymb,amsmath,amsthm}
\usepackage{proof}
\usepackage{graphics} 

\newtheorem{theorem}{Theorem}
\newtheorem{lemma}[theorem]{Lemma}

\newcommand{\ie}{{\em i.e.}\xspace}
\newcommand{\eg}{{\em e.g.}\xspace}

\newcommand{\etal}{{\em et al.}\xspace}

\newcommand{\G}{\ensuremath{\cal G}\xspace}
\newcommand{\Gm}{\ensuremath{\cal G^-}\xspace}
\newcommand{\new}{\reflectbox{\ensuremath{\mathsf{N}}}}
\newcommand{\Lra}{\Longrightarrow}
\newcommand{\lra}{\longrightarrow}
\newcommand{\supp}{{\rm supp}}
\newcommand{\overvec}{\overrightarrow}

\newcommand{\defR}{\hbox{\sl def}\mathcal{R}}
\newcommand{\eqR}{=\mathcal{R}}
\newcommand{\topR}{\top\mathcal{R}}
\newcommand{\lorR}{\lor\mathcal{R}}
\newcommand{\landR}{\land\mathcal{R}}
\newcommand{\nablaR}{\nabla\mathcal{R}}
\newcommand{\existsR}{\exists\mathcal{R}}

\newcommand{\name}[1]{\hbox{\sl name} \; #1}
\newcommand{\fresh}[2]{\hbox{\sl fresh} \; #1 \; #2}
\newcommand{\abst}[3]{\hbox{\sl abst} \; #1 \; #2 \; #3}
\newcommand{\swap}[4]{\hbox{\sl swap} \; #1 \; #2 \; #3 \; #4}

\newcommand{\seq}[2]{\hbox{\sl seq} \; #1 \; #2}
\newcommand{\prog}[2]{\hbox{\sl prog} \; #1 \; #2}
\newcommand{\member}[2]{\hbox{\sl member} \; #1 \; #2}

\newcommand{\ent}{\mathrel{{:}-}}

\newcommand{\ignore}[1]{}

\begin{document}

\conferenceinfo{PPDP'10,}{July 26--28, 2010, Hagenberg, Austria.}
\CopyrightYear{2010}
\copyrightdata{978-1-4503-0132-9/10/07}

\title{Relating Nominal and Higher-order Abstract Syntax
  Specifications}
\authorinfo
 {Andrew Gacek\titlenote{
     This work has been supported by INRIA through the ``Equipes
     Associ{\'e}es'' Slimmer and by the NSF Grant CCF-0917140.
     Opinions, findings, and conclusions or recommendations expressed
     in this papers are those of the author and do not necessarily
     reflect the views of the National Science Foundation.
}}
 {INRIA Saclay -- \^Ile-de-France \& LIX/\'Ecole polytechnique \\
  Palaiseau, France}
 {gacek@lix.polytechnique.fr}
\maketitle

\begin{abstract}
Nominal abstract syntax and higher-order abstract syntax provide a
means for describing binding structure which is higher-level than
traditional techniques. These approaches have spawned two different
communities which have developed along similar lines but with subtle
differences that make them difficult to relate. The nominal abstract
syntax community has devices like names, freshness, name-abstractions
with variable capture, and the $\new$-quantifier, whereas the
higher-order abstract syntax community has devices like
$\lambda$-binders, $\lambda$-conversion, raising, and the
$\nabla$-quantifier. This paper aims to unify these communities and
provide a concrete correspondence between their different devices. In
particular, we develop a semantics-preserving translation from
$\alpha$Prolog, a nominal abstract syntax based logic programming
language, to \Gm, a higher-order abstract syntax based logic
programming language. We also discuss
higher-order judgments, a common and powerful tool for specifications
with higher-order abstract syntax, and we show how
these can be incorporated into \Gm. 
This establishes \Gm as a language
with the power of higher-order abstract syntax, the fine-grained
variable control of nominal specifications, and the desirable
properties of higher-order judgments.
\end{abstract}




\section{Introduction}
\label{sec:intro}

Many approaches and languages have been proposed for encoding logical
specifications of systems with binding. One popular approach is based
on {\em nominal logic} which formalizes a notion of
$\alpha$-equivalence classes along with related devices
\cite{pitts03ic}. This has led to the $\alpha$Prolog language which
allows for executing specifications based on nominal logic
\cite{cheney04iclp}. Another popular approach is based on {\em
  higher-order abstract syntax} which uses a weak $\lambda$-calculus
to represent binding in object systems \cite{miller87slp,
  pfenning88pldi}. Formalizing specification based on higher-order
abstract syntax requires a framework with devices for manipulating and
forming judgments over $\lambda$-terms. The most notable examples of
such frameworks are LF \cite{harper93jacm} and $\lambda$Prolog
\cite{nadathur88iclp} which use higher-order techniques for
representing both syntax and judgments.

The success of both the nominal and higher-order approaches has lead
to questions regarding their relationship and relative merits such as
naturalness and expressiveness. Higher-order abstract syntax provides
a high-level treatment of binding and is often used with higher-order
judgments to produce elegant specifications. These specifications
benefit from a free notion of substitution inherited from the
specification language and from nice properties which can be used when
reasoning, \eg, that substitution for free variables preserves the
validity of judgments. On the other hand, nominal approaches require
substitution issues to be dealt with manually, but allow object
variables to be manipulated directly. This results in natural
specifications when fine-grained control over object variables is
required. The same naturalness is not found in similar situations when
using higher-order judgments. This is not a limitation of the
high-level treatment of binding provided by higher-order abstract
syntax, but rather of the companion notion of higher-order judgments.
In fact, in this paper we show that higher-order abstract syntax in a
suitable framework is capable of at least the same naturalness and
expressiveness as nominal logic specifications. We do this by
developing and proving correct a direct translation from
$\alpha$Prolog programs to definitions in \Gm, a logic with
higher-order abstract syntax.

Higher-order judgments play an important role in higher-order abstract
syntax specifications because of their elegance and nice properties.
It is disappointing that they are abandoned in order to make this
connection between nominal and higher-order abstract syntax
specifications. We show, however, that higher-order judgments can be
encoded in \Gm so that their nice features are preserved. Thus \Gm is
a language in which the benefits of both nominal logic and
higher-order judgments can be realized.

Let us consider an example to demonstrate the already close
correspondence between specifications based on nominal and
higher-order abstract syntax. The following $\alpha$Prolog program
describes type checking for $\lambda$-terms.
\begin{align*}
&\forall G,X,T. [tc(G,var(X),T) \ent lookup(X,T,G)] \\
&\forall G,E_1,E_2,T'. [tc(G,app(E_1,E_2), T') \ent \\
&\hspace{2.5cm} \exists T. tc(G,E_1,arr(T,T')) \land tc(G,E_2,T)] \\
&\new x.\forall G, E, T, T'.
  [tc(G, lam(\langle x\rangle E), arr(T, T')) \ent \\
&\hspace{2.5cm}
    x \# G \land tc(bind(x,T,G), E, T')]
\end{align*}
The last clause illustrates the specification of binding structure
and features the nominal $\new$-quantifier for fresh variable
names, name-abstraction $\langle x\rangle E$ for denoting object
binding structure, and the fresh relation $x \# G$ for enforcing a
freshness side-condition. The same program can be specified in \Gm
using the following definitional clauses.
\begin{align*}
&\forall G, X, T. [tc\ G\ (var\ X)\ T \triangleq lookup\ X\ T\ G] \\
&\forall G, E_1, E_2, T'. [tc\ G\ (app\ E_1\ E_2)\ T' \triangleq \\
&\hspace{2.5cm}
   \exists T. tc\ G\ E_1\ (arr\ T\ T') \land tc\ G\ E_2\ T] \\
&\forall G, E, T, T'.
  [tc\ G\ (lam\ \lambda x. E\, x)\ (arr\ T\ T') \triangleq \\
&\hspace{2.5cm} \nabla x. tc\ (bind\ x\ T\ G)\ (E\, x)\ T']
\end{align*}
The last clause here features the $\nabla$-quantifier for fresh
variable names, a $\lambda$-binder for denoting object binding
structure, and a function application $(E\, x)$ denoting a
substitution. In addition, the variable quantification order in the
last clause enforces the freshness side-condition: since $x$ is
quantified inside the scope of $G$, no instantiation for the latter can
contain the former. The translation we present in this paper actually
generates these definitional clauses given the original
$\alpha$Prolog program. By studying this translation and proving it
correct, we can pin down the exact relationship between the nominal
and higher-order devices of these two specifications.

It is important to note that this paper is not an attempt to argue
that one approach or another is irrelevant. Nominal techniques
embed nicely in existing theorem provers \cite{urban08jar}, and
higher-order techniques enable high-level specification and
reasoning \cite{gacek08lfmtp, harper07jfp}. In addition, this
paper does not attempt to relate implementation issues associated
with executing nominal and higher-order abstract syntax
specifications, such as higher-order, nominal, or equivariant
unification. Such relationships have been investigated in other works
\cite{cheney05unif, urban04tcs}.

The paper is organized as follows. We describe $\alpha$Prolog in
Section~\ref{sec:alpha-prolog}, \Gm in Section~\ref{sec:g}, and
the translation in Section~\ref{sec:translation}. We discuss the
relationship with higher-order judgments in Section~\ref{sec:hoj},
and we conclude in Section~\ref{sec:conc}.

\section{$\alpha$Prolog}
\label{sec:alpha-prolog}

The syntax of $\alpha$Prolog is made up of terms, goals, and program
clauses which are defined by the following grammars, respectively:
\begin{align*}
t,u ::= &\ a \mid X \mid f(\vec{t}) \mid (a\ b)\cdot t \mid \langle a
\rangle t \\
G ::= &\ \top \mid p(\vec{t}) \mid a \# t \mid t \approx u \mid \\
&\ G \land G' \mid G \lor G' \mid \exists X.G \mid \new a. G \\
D ::= &\ \new \vec{a}. \forall \vec{X}.
  [p(\vec{t}) \ent G]
\end{align*}

For terms, $a$ and $b$ denote {\em names} which are used to
represent object language variables, $X$ denotes a first-order
variable, and $f$ denotes a function symbol. Constants are encoded
as function symbols which take no arguments. The construct $(a\
b)\cdot t$ denotes a {\em swapping} of the names $a$ and $b$
within the term $t$. The construct $\langle a \rangle t$ is called
a {\em name-abstraction} and is used to represent object language
bindings. We assume that all terms are well-typed
according to a monomorphic typing discipline. We will avoid the
details of typing, except to note that $\alpha$Prolog requires
names to belong to distinguished {\em name types} which are not
inhabited by any other terms.

Goals are constructed from the usual logical connectives. The goal
$a \# t$ is a {\em freshness} constraint and holds when the name
$a$ does not occur free relative to name-abstractions in $t$.
The equality goal $t \approx u$ denotes
a notion of $\alpha$-convertibility which treats name-abstraction
as a binder. The goal $\new a.G$ represents a binding for the name
$a$ in the scope of $G$.

We assume a single form for program clauses. While $\alpha$Prolog
admits richer forms, these are normalizable to the one presented here,
possibly by inserting freshness constraints. The expression
$p(\vec{t})$ is called the head of the clause. A predicate can appear
in the head of multiple clauses. A clause is well-formed if it
contains no free variables or free names. Note that the name $a$
appears free in $\langle a\rangle a$ but not in $\new a.G$ since the
former is not a real binder while the latter is. We shall consider
only well-formed program clauses from here onwards. An $\alpha$Prolog
program is a set of program clauses.

We have presented the {\em name-restricted} subset of
$\alpha$Prolog where $a$ and $b$ must be names in
$(a\ b)\cdot t$, $\langle a\rangle t$, and $a \# t$. We will focus
on this subset for most of the paper, but will eventually lift
this restriction and treat full $\alpha$Prolog.

\begin{figure}[t]
\begin{align*}
&(a\ b) \cdot a = b \\
&(a\ b) \cdot b = a \\
&(a\ b) \cdot a' = a' \;\;\;\; (a \neq a' \neq b) \\
&(a\ b) \cdot f(\vec{t}) = f(\overvec{(a\ b)\cdot t}) \\
&(a\ b) \cdot \langle a'\rangle t = \langle (a\ b)\cdot a' \rangle
  (a\ b) \cdot t
\end{align*}
\vspace{0.1cm}
\begin{equation*}
\begin{array}{cc}
\infer{\models a \# b}{a \neq b} &
\infer{\models a \# f(\vec{t})}
      {\models a \# t_1 & \ldots & \models a\# t_n}
\\[10pt]
\infer{\models a \# \langle a \rangle t}{} &
\infer{\models a \# \langle b \rangle t}
       {\models a \# b & \models a \# t}
\end{array}
\end{equation*}
\vspace{0.1cm}
\begin{equation*}
\begin{array}{cc}
\infer{\models a \approx a}{} &
\infer{\models f(\vec{t}) \approx f(\vec{u})}
      {\models t_1 \approx u_1 & \ldots & \models t_n \approx u_n}
\\[10pt]
\infer{\models \langle a\rangle t \approx \langle a \rangle u}
      {\models t \approx u} &
\infer{\models \langle a\rangle t \approx \langle b \rangle u}
      {\models a \# u & \models t \approx (a\ b)\cdot u}
\end{array}
\end{equation*}
\caption{Swapping, freshness, and equality for ground nominal terms}
\label{fig:basics}
\end{figure}

An $\alpha$Prolog expression is a term, list of terms, goal, or
program clause. An expression is {\em ground} if it does not contain
any free variables, though it may contain free names.
We define the meaning of swapping, freshness, and equality for
ground nominal terms as shown in Figure~\ref{fig:basics}. We
extend the notion of swapping to goals in the expected way with
$(a\ b)\cdot \new a'. G = \new a'. (a\ b)\cdot G$ where $a \neq a'
\neq b$ and with $(a\ b)\cdot \exists X.G = \exists X.(a\ b)\cdot
G$. To make sense of this last equation, we define $(a\ b)\cdot X
= X$, though we leave this out of the formal definition of
swapping since we intend to focus on ground terms and goals. We
define a permutation $\pi$ as a composition of zero or more
swappings and we write $\pi.e$ to denote the effect of applying
the swappings in $\pi$ to the expression $e$.

We assume the standard notions of binding for quantifiers and use
$e[t/X]$ to denote capture-avoiding substitution of the term $t$
for the variable $X$ in the expression $e$. Similarly we write
$\theta$ for a simultaneous substitution for zero or more
variables and $e\theta$ for its application to the expression $e$.
Note that name-abstractions are not really binders and thus
substitution can cause name capture, \eg, $(\langle a\rangle
X)[a/X] = \langle a\rangle a$.

We view computation in $\alpha$Prolog as the search for a proof of
the sequent $\Delta \Lra G$ where $\Delta$ is a set of program
clauses and $G$ is a goal. A sequent is well-formed if $G$ is ground,
and we shall consider only well-formed sequents from here onwards. Our
view of $\alpha$Prolog purposefully ignores issues related to an
actual implementation such as searching for instantiations for
existentially quantified variables and related issues of
unification \cite{cheney08toplas, urban04tcs}.

The proof rules for $\alpha$Prolog are shown in
Figure~\ref{fig:rules}. In the {\sc exists} rule, $t$ may contain
any names and similarly for the substitution $\theta$ in {\sc
backchain}. In both rules, the substitutions cannot contain free
variables, thereby ensuring that goals remain ground during proof
search. In the {\sc backchain} rule we use the relation $\approx$
between lists of terms to mean that respective terms in the two
list satisfy the $\approx$ relation.

\begin{figure}[t]
\begin{align*}
\infer[\mbox{\sc true}]
      {\Delta \Lra \top}
      {}
&&
\infer[\mbox{\sc fresh}]
      {\Delta \Lra a \# t}
      {\models a \# t}
&&
\infer[\mbox{\sc equal}]
      {\Delta \Lra t \approx u}
      {\models t \approx u}
\end{align*}
\begin{align*}
\infer[\mbox{\sc and}]
      {\Delta \Lra G_1 \land G_2}
      {\Delta \Lra G_1 & \Delta \Lra G_2}
&&
\infer[\mbox{\sc or}]
      {\Delta \Lra G_1 \lor G_2}
      {\Delta \Lra G_i}
\end{align*}
\begin{align*}
\infer[\mbox{\sc exists}]
      {\Delta \Lra \exists X . G}
      {\Delta \Lra G[t/X]} &&
\infer[\mbox{\sc new}]
      {\Delta \Lra \new a . G}
      {\Delta \Lra G}
\end{align*}
\begin{align*}
\infer[\mbox{\sc backchain}]
      {\Delta \Lra p(\vec{t})}
      {\Delta \Lra \pi.(G\theta)}
\end{align*}
\begin{center}
Where $\new\vec{a}.\forall\vec{X}.[p(\vec{u}) \ent G] \in
  \Delta$ and $\pi$ is a permutation and $\theta$ is a
  substitution for $\vec{X}$ such that $\vec{t} \approx
  \pi.(\vec{u}\theta)$.
\end{center}
\caption{Proof rules for $\alpha$Prolog}
\label{fig:rules}
\end{figure}

As an example, let $\Delta$ be the set of program clauses for type
checking given in the introduction and the assumed clauses for the
$lookup$ predicate. The object term $\lambda z.\lambda z.z$ can be
assigned the type $\alpha \to \beta \to \beta$ for any types
$\alpha$ and $\beta$. The corresponding derivation for this is
shown in Figure~\ref{fig:ex-alpha-prolog}. On the other hand, the
term cannot be assigned the type $\alpha \to \beta \to \alpha$
when $\alpha$ is not equal to $\beta$. To do so would require the
derivation to use the same name for both the first and second
abstractions in the term. This is disallowed by the use of the
freshness predicate.

\begin{figure*}[t]
\begin{equation*}
\infer
 [\mbox{\sc backchain}]
 {\Delta \Lra tc(nil, lam(\langle a\rangle lam(
\langle a\rangle var(a))), arr(\alpha,arr(\beta,\beta)))}
 {\infer
  [\mbox{\sc and}]
  {\Delta \Lra a \# nil \land
   tc(bind(a,\alpha,nil), lam(\langle a\rangle var(a)), arr(\beta,\beta))}
  {\infer[\mbox{\sc fresh}]{\Delta \Lra a \# nil}{\models a \# nil} &
   \infer[\mbox{\sc backchain}]
    {\Delta \Lra tc(bind(a,\alpha,nil), lam(\langle a\rangle
       var(a)), arr(\beta,\beta))}
    {\infer[\mbox{\sc and}]
      {\Delta \Lra b \# bind(a,\alpha,nil) \land
       tc(bind(b,\beta,bind(a,\alpha,nil)), var(b), \beta)}
      {\infer[\mbox{\sc fresh}]
        {\Delta \Lra b \# bind(a,\alpha,nil)}
        {\models b \# bind(a,\alpha,nil)}
       &
       \infer[\mbox{\sc backchain}]
        {\Delta \Lra tc(bind(b,\beta,bind(a,\alpha,nil)),
         var(b), \beta)}
        {\deduce
          {\Delta \Lra
           lookup(b,\beta,bind(b,\beta,bind(a,\alpha,nil)))}
          {\vdots}
        }
      }
    }
  }
 }
\end{equation*}
\caption{A derivation of $tc$ in $\alpha$Prolog}
\label{fig:ex-alpha-prolog}
\end{figure*}

An important characteristic of $\alpha$Prolog derivations is that
they are {\em equivariant}, \ie, unchanged by permutations of
names. This property ensures that the particular choice of names
used in a derivation is immaterial. More formally, one can
inductively define a notion of applying a permutation to a
derivation so that its structure and correctness are preserved.
Assuming this, we will treat as equivalent those derivations which
differ only by a permutation of names.

Cheney and Urban \cite{cheney08toplas} introduce a Herbrand model
based semantics for nominal logic which we can use to show the
relative consistency and completeness of our presentation of
$\alpha$Prolog. In particular, given a set of nominal logic
formulas $\Gamma$ and a nominal logic formula $\phi$, they write
$\Gamma \models \phi$ to indicate that any Herbrand model for all
the elements of $\Gamma$ is a model of $\phi$. Using this notion,
we can prove the following.
\begin{theorem}
Let $\Delta$ be a set of program clauses and $G$ a ground goal.
Then
$\Delta \models G$ holds if and only if $\Delta \Lra G$ has a
proof.
\end{theorem}
\begin{proof}
The forwards direction uses the fact that a least Herbrand model
exists for $\Delta$. The backwards direction is by induction on
the derivation of $\Delta \Lra G$.
\end{proof}

\section{The Logic \Gm}
\label{sec:g}

\begin{figure}[t]
\begin{align*}
\infer[\topR]
      {\lra \top}
      {}
&&
\infer[\eqR]
      {\lra t = t}
      {}
\end{align*}
\begin{align*}
\infer[\landR]
      {\lra B_1 \land B_2}
      {\lra B_1 & \lra B_2}
&&
\infer[\lorR]
      {\lra B_1 \lor B_2}
      {\lra B_i}
\end{align*}
\begin{align*}
\infer[\existsR]
      {\lra \exists x . B}
      {\lra B[t/x]}
&&
\infer[\nablaR,\ a\notin\supp(B)]
      {\lra \nabla x . B}
      {\lra B[a/x]}
\end{align*}
\begin{align*}
\infer[\defR]
      {\lra p\ \vec{t}}
      {\lra B\theta}
\end{align*}
\begin{center}
Where $\forall\vec{x}. [(\nabla\vec{z}. p\ \vec{u}) \triangleq B]
\in {\cal D}$ and $\theta$ is a substitution for $\vec{z}$ and
$\vec{x}$ such that each $z_i\theta$ is a unique nominal constant,
$\supp(\vec{x}\theta) \cap \{\vec{z}\theta\} = \emptyset$, and
$\vec{t} = \vec{u}\theta$.
\end{center}
\caption{Proof rules for \Gm}
\label{fig:g-rules}
\end{figure}

The logic \Gm is a first-order logic over a higher-order term language
where specifications are encoded as fixed-point definitions for
predicates. This is in contrast to languages like $\alpha$Prolog and
$\lambda$Prolog which use Horn-like clauses to encode specifications.
The reason for this departure is that \Gm is actually a subset of a
richer logic \G which is designed for reasoning \cite{gacek08lfmtp,
  gacek08lics}, and
within this larger setting, mechanisms like case-analysis and
induction give a different meaning to Horn-like clauses than in pure
specification logics. Our presentation of \Gm differs
from that of \G in some regards, but it is still a proper
subset and can be shown to be sound relative to \G.

The syntax of terms in \Gm is as follows:
\begin{align*}
t,u &::= x \mid c \mid a \mid (t\, u) \mid \lambda x. t
\end{align*}
Here $x$ denotes a variable, $c$ denotes a constant, and $a$
denotes a {\em nominal constant}. The term $\lambda x. t$ denotes
a binding for the variable $x$ in the scope of $t$, and we
assume the corresponding standard notations of free and bound
variables and capture-avoiding substitution. The scope of a
$\lambda$ is as far to the right as possible. We will assume that
all terms are in $\beta\eta$-long form and that all comparisons
between terms are relative to the standard rules of
$\lambda$-conversion. Application associates to the left and we
write $p\ \vec{t}$ to abbreviate $p\ t_1\ \cdots\ t_n$. We
restrict our attention to terms which are well-typed relative to a
monomorphic typing system. We will avoid the
details of typing, except to note that \Gm allows nominal constants
only at pre-designated {\em nominal types} which may or may not be
inhabited by other terms.

Formulas in \Gm are terms of a distinguished type $o$. We
introduce the constant $\top$ of type $o$ and the infix constants
$\land$ and $\lor$ of type $o \to o \to o$. For each type $\tau$
that does not contain $o$ we include the equality constant
$=_\tau$ of type $\tau \to \tau \to o$ and the constants
$\exists_\tau$ and $\nabla_\tau$ of type $(\tau \to o) \to o$. We
place the further restriction on $\nabla_\tau$ that $\tau$ must be a
nominal type. We
drop subscripts when they can be inferred from the context. We
abbreviate the formulas $\exists (\lambda x.t)$ and $\nabla
(\lambda x.t)$ as $\exists x.t$ and $\nabla x.t$, respectively. In
summary, the formulas of \Gm are described by the following
grammar.
\begin{align*}
B, C &::= \top \mid p\ \vec{t} \mid t = u \mid B \land C \mid B
\lor C \mid \exists x.B \mid \nabla z.B
\end{align*}
Here $p$ denotes any additional predicate symbol, \ie, constant of
type $\tau_1 \to \ldots \to \tau_n \to o$.

We use $\nabla$ to quantify over fresh nominal constants. The
treatment of $\nabla$ in \Gm is based on the so-called {\em
nominal} $\nabla$-quantifier \cite{tiu06lfmtp} rather than the
earlier {\em minimal} $\nabla$-quantifier \cite{miller05tocl}. The
essential difference is that the nominal $\nabla$-quantifier
admits exchange, $\nabla x.\nabla y.B \equiv \nabla y.\nabla x.B$,
and weakening and strengthening, $\nabla x.B \equiv B$ if $x$ does
not appear in $B$. We prefer the nominal treatment since nominal
constants are often used to represent variable names, and these
equivalences match our intuitions about fresh variable names. In
addition, the nominal treatment often results in simplified
meta-theory and reasoning.

We define the support of a term as the nominal constants which
appear in it:
\begin{align*}
\supp(x) &= \supp(c) = \emptyset &
\supp(a) &= \{a\} \\
\supp(t\ u) &= \supp(t) \cup \supp(u) &
\supp(\lambda x. t) &= \supp(t)
\end{align*}
We define the support of a list of terms as the union of their
supports.

Specifications are realized in \Gm through {\em fixed-point
  definitions}. Fixed-point definitions are given by a set of {\em
  definitional clauses}, each of the following form:
\begin{equation*}
\forall \vec{x}. [(\nabla\vec{z} . p\ \vec{t}) \triangleq B]
\end{equation*}
Here $\nabla\vec{z} . p\ \vec{t}$ and $B$ must be formulas with
empty support and free variables only among $\vec{x}$. The formula
$\nabla\vec{z} . p\ \vec{t}$ is called the head of the clause. A
predicate symbol $p$ may appear in the head of multiple clauses.
The logic \Gm is parametrized by a set of definitional clauses
which we will call $\cal D$.

We view computation in \Gm as the search for a proof of the sequent
$\lra B$ where $B$ is a closed formula which may contain nominal
constants. The proof rules for \Gm are presented in
Figure~\ref{fig:g-rules}. In the $\existsR$ rule we assume $t$ is a
closed term which may contain any nominal constants and similar for
the substitution in the $\defR$ rule. In the $\defR$ rule we assume a
notion of pairwise equality on lists of terms. We require the support
of $\vec{x}\theta$ to be disjoint from $\vec{z}\theta$ to reflect the
order of quantifiers in the definitional clause. In the vocabulary of
Miller \etal \cite{miller91apal}, the rules of \Gm allow only uniform
proofs and thus \Gm is an abstract logic programming language.

Assuming the definition of $tc$ from the introduction and a suitable
definition of $lookup$,
Figure~\ref{fig:ex-g} shows that $\lambda z. \lambda z. z$ can
be assigned the type $\alpha \to \beta \to \beta$. Note that it is
not possible to assign the type $\alpha \to \beta \to \alpha$ when
$\alpha$ is not equal to $\beta$ due to side-condition on the
$\nablaR$ rule.

\begin{figure}[t]
\begin{equation*}
\infer
 [\defR]
 {\lra tc\ nil\ (lam\ \lambda z. lam\ \lambda z. var\ z)\ 
    (arr\ \alpha\ (arr\ \beta\ \beta))}
 {\infer
  [\nablaR]
  {\lra \nabla x. tc\ (bind\ x\ \alpha\ nil)\
    (lam\ \lambda z. var\ z)\ (arr\ \beta\ \beta)}
  {\infer[\defR]
   {\lra tc\ (bind\ a\ \alpha\ nil)\ (lam\ \lambda z. var\ z)\ 
      (arr\ \beta\ \beta)}
   {\infer[\nablaR]
    {\lra \nabla x. tc\ (bind\ x\ \beta\ (bind\ a\ \alpha\ nil))\ 
      (var\ x)\ \beta}
    {\infer[\defR]
     {\lra tc\ (bind\ b\ \beta\ (bind\ a\ \alpha\ nil))\ 
      (var\ b)\ \beta}
     {\deduce
       {\lra lookup\ b\ \beta\ 
        (bind\ b\ \beta\ (bind\ a\ \alpha\ nil))}
      {\vdots}
     }
    }
   }
  }
 }
\end{equation*}
\caption{A derivation of $tc$ in \Gm}
\label{fig:ex-g}
\end{figure}

Derivations in \Gm are equivariant with respect to nominal
constants, \ie, the particular nominal constants used in a \Gm
derivation are irrelevant. Given a permutation of nominal
constants, one can inductively define a notion of applying that
permutation to a derivation so that the structure and correctness
are preserved. Thus we will treat as equivalent those derivations
which differ only by a permutation of nominal constants.

\section{The Translation}
\label{sec:translation}

\begin{figure*}[t]
\begin{align*}
\phi(a) &= a
&
\phi(X\, \vec{a}) &= X\, \vec{a}
&
\phi((a\ b)\cdot t) &= (a\ b)\cdot \phi(t)
&
\phi(f(\vec{t})) &= f\ \overvec{\phi(t)}
&
\phi(\langle a\rangle t) &= \lambda a. \phi(t)
\end{align*}
\smallskip
\begin{align*}
\phi_{\vec{a}}\left(p\ \vec{t}\right) &=
  \nabla\vec{a}. p\ \overvec{\phi(t)}
&
\phi_{\vec{a}}\left(G_1\land G_2\right) &=
  \phi_{\vec{a}}(G_1) \land \phi_{\vec{a}}(G_2)
\\
\phi_{\vec{a}}\left(\top\right) &=
  \top
&
\phi_{\vec{a}}\left(G_1\lor G_2\right) &=
  \phi_{\vec{a}}(G_1) \lor \phi_{\vec{a}}(G_2)
\\
\phi_{\vec{a}}\left(a\#t\right) &=
  \nabla\vec{a}. \fresh {\phi(a)} {\phi(t)}
&
\phi_{\vec{a}}\left(\exists X.G\right) &=
  \exists X.\phi_{\vec{a}}(G[X\, \vec{a}/X])
\\
\phi_{\vec{a}}\left(t \approx u\right) &=
  \nabla \vec{a}. (\phi(t) = \phi(u))
&
\phi_{\vec{a}}\left(\new b.G\right) &=
  \phi_{\vec{a}b}(G)
\end{align*}
\smallskip
\begin{align*}
\phi\left(\new\vec{a}. \forall \vec{X}. [p(\vec{t}) \ent
  G]\right) &=
\forall\vec{X}.[(\nabla\vec{a}. p\ \overvec{\phi(t\sigma)})
  \triangleq \phi_{\vec{a}}(G\sigma))] \;\;
\mbox{where $\sigma = \{ X\, \vec{a}/X \mid X \in \vec{X} \}$}
\end{align*}
\caption{Translation from $\alpha$Prolog to $\Gm$}
\label{fig:translation}
\end{figure*}

Looking at the rules for $\alpha$Prolog and \Gm we can already see
a strong similarity. In large part, this is because we have
developed a view of $\alpha$Prolog free from implementation
details and have carved out \Gm from the richer logic of \G. We
have, however, remained faithful to both languages.

One might expect a very simple translation from $\alpha$Prolog to
\Gm which maps $\new$ to $\nabla$, $\approx$ to $=$, names to
nominal constants, and name-abstraction to $\lambda$-abstraction.
This is not far from the truth, but there is an important nuance
concerning the treatment of abstractions in the two systems. In
essence, the name-abstraction of $\alpha$Prolog allows names to be
captured during substitution while the $\lambda$-abstraction
requires capture-avoiding substitution. For example, consider the
$\alpha$Prolog goal $\new a.\exists X. (\langle a\rangle X \approx
\langle b\rangle b)$. This goal is provable using {\sc new} and
{\sc exists} with $X$ as $a$ thus yielding $\langle a\rangle a
\approx \langle b\rangle b$ which is true. Now, a naive and
incorrect translation of the original goal into \Gm might produce
$\nabla a.\exists X. (\lambda a.X = \lambda b.b)$. Notice that the
two occurrences of $a$ in this goal represent distinct binders and
thus the goal is equivalent to $\nabla y.\exists X. (\lambda z.X =
\lambda b.b)$. This formula is not provable in \Gm since
capture-avoiding substitution does not allow any value for $X$ to
be captured by the binder for $z$. Instead, we need a translation
which makes the possible variable captures in $\alpha$Prolog
explicit. For instance, the original goal may be translated to
essentially $\exists X.(\lambda a. X\, a = \lambda b.b)$ which has
the solution $X = \lambda z.z$. In this formula, the variable $X$
has been {\em raised} over $a$ to indicate its possible dependence
on it. This is a standard technique which is used in relating nominal
and higher-order term languages \cite{urban04tcs}. In the actual
translation we will use raising to encode all  such dependencies.

The translation from $\alpha$Prolog to \Gm is presented in
Figure~\ref{fig:translation} and makes use of some new notation
which we define now.

In the translation for terms, we map names to nominal constants,
and for simplicity we overload notation to use the same names for
both. We use the same overloading for bound variables and function
symbols. We abuse notation in our translation to allow bound
variables in $\alpha$Prolog to be raised over nominal constants.
This is just an intermediate form which is translated to a \Gm
bound variable with the same name raised over the same nominal
constants. The translation for swappings produces a similar
operation applied to a $\Gm$ term which we represent with the same
notation. The meaning of a swapping applied to a \Gm term is to
replace all occurrences of one nominal constant with another and
vice-versa. In contrast to $\alpha$Prolog, this operation can be
carried out completely even for non-ground terms since all
variables of the translation are raised over the existing nominal
constants they may depend on. Thus, swapping can be carried out on
the nominal constants over which variables are raised without
having to know the eventual value of such variables. Lastly, the
translation for name-abstractions maps them to $\lambda$-binders.
Although nominal constants and bound variables are from separate
syntactic classes in \Gm, we abuse notation here and in the future
to write a binder for a nominal constant. The meaning of $\lambda
a.t$ where $a$ is a nominal constant is $\lambda x.t'$ where $x$
is a fresh bound variable name and $t'$ is the result of replacing
all occurrences of $a$ in $t$ with $x$.

As indicated in the initial discussion, our translation needs to push
$\nabla$-binders underneath $\exists$-binders so that the dependencies
can be made explicit. This is embodied in our translation for goals
which is parametrized by a list of names which correspond to
$\nabla$-bound variables being pushed down to the atomic formulas. The
following equivalences describe how the $\nabla$-quantifier can be
pushed down in a formula.
\begin{align*}
\nabla x.\top &\equiv \top \\
\nabla x.(B \land C) &\equiv (\nabla x.B) \land (\nabla x.C) \\
\nabla x.(B \lor C) &\equiv (\nabla x.B) \lor (\nabla x.C) \\
\nabla x.\exists X.B &\equiv \exists X.\nabla x.B[X\, x/X]
\end{align*}
In the translation for a freshness goal we make use of a distinguished
predicate {\sl fresh} which we assume is defined by the single
definitional clause $\forall x. (\nabla z. \fresh z x) \triangleq
\top$. Thus $\lra \fresh a t$ is provable if and only if $a$ is a
nominal constant which does not appear in $t$. Lastly, note that we
translate the $\new$-quantifier to $\nabla$-quantifier by adding the
quantified name to the list of eventually $\nabla$-bound variables.

The translation for program clauses embodies essentially the same
ideas as for translating goal formulas. The outer
$\new$-quantifiers are translated to $\nabla$-quantifiers that
need to be pushed underneath the universal quantifiers. This
exchange induces the same raising substitution as when pushing
$\nabla$-quantifiers underneath existential quantifiers. When
$\Delta$ is a set of program clauses we define $\phi(\Delta) =
\{\phi(D) \mid D \in \Delta\}$.

Note that aside from changing the scope of $\nabla$-quantifiers,
the translation essentially preserves term and logic structure.
Moreover, the translation makes the expected connections between
$\new$ and $\nabla$, $\approx$ and $=$, names and nominal
constants, and name-abstraction and $\lambda$-abstraction.

\subsection{Examples}

We now present a few examples to illustrate the translation and
also to suggest some simple ways in which the results may be
improved. In presenting $\alpha$Prolog program clauses we will
elide outermost $\new$ and $\forall$-quantifiers and instead use
the convention that all free lowercase symbols denote names and
all free uppercase symbols denote variables, all of which are
captured by program clause quantifiers. For \Gm definitions we
elide the outermost $\forall$-quantifiers and assume that all
capitalized symbols denote such universally quantified variables.
These examples are taken from Cheney and Urban
\cite{cheney08toplas}.

\subsubsection{Type checking}

Consider again the example from the introduction, where the
following program clauses specify type checking for $\lambda$-terms.
\begin{align*}
&tc(G,var(X),T) \ent lookup(X,T,G) \\
&tc(G,app(E_1,E_2), T') \ent \\
&\hspace{1cm}    \exists T. tc(G,E_1,arr(T,T')) \land tc(G,E_2,T) \\
&tc(G, lam(\langle x\rangle E), arr(T, T')) \ent \\
&\hspace{1cm}
    x \# G \land tc(bind(x,T,G), E, T')
\end{align*}
Here we assume $lookup$ is defined in the expected way.
These program clauses translate to the following definitional clauses.
\begin{align*}
&tc\ G\ (var\ X)\ T \triangleq lookup\ X\ T\ G \\
&tc\ G\ (app\ E_1\ E_2)\ T' \triangleq \\
&\hspace{1cm}
   \exists T. tc\ G\ E_1\ (arr\ T\ T') \land tc\ G\ E_2\ T \\
&(\nabla x. tc\ (G\, x)\ (lam\ \lambda x. E\, x)\ (arr\ (T\, x)\ (T'\, x)))
     \triangleq \\
&\hspace{1cm}
   (\nabla x. \fresh x (G\, x)) \land~ \\
&\hspace{1cm}
   (\nabla x. tc\ (bind\ x\ (T\, x)\ (G\, x))\ (E\, x)\ (T'\, x))
\end{align*}
In general, a few simplifications can improve the results of
the translation. We illustrate these here as applied to the last
clause for $tc$. First, by
examining types we can recognize that the object types $T$ and
$T'$ cannot actually depend on the object term variable $x$ and
thus they do not need to be raised over it. More formally, we can
use a notion like {\em subordination} to detect such vacuous
dependencies \cite{virga99phd}. Second, the freshness constraint
on $G$ can be solved statically: we know that $G$ must not depend
on its first argument. Finally, after performing the previous two
simplifications we can recognize that the $\nabla$-quantifier in
the head of the definition is vacuous and can thus be dropped. In
the end we are left with the following definitional clause:
\begin{align*}
&tc\ G\ (lam\ \lambda x. E\, x)\ (arr\ T\ T') \triangleq \\
&\hspace{1cm} \nabla x. tc\ (bind\ x\ T\ G)\ (E\, x)\ T'
\end{align*}
The definitional clauses for $tc$ now match what one would expect to
write in \Gm. Indeed, they are exactly specification presented in the
introduction.

\subsubsection{Polymorphic type generalization}

The following program clauses describe a relationship among a
polymorphic type, a list of distinct names for the binders in that
type, and
the resulting monomorphic type which comes from substituting the names
for the binders.
\begin{align*}
&spec(monoTy(T),nil,T) \ent \top \\
&spec(polyTy(\langle a\rangle P), cons(a,L), T) \ent \\
&\hspace{1cm} a \# L \land spec(P,L,T)
\end{align*}
This is translated to the following definitional clauses:
\begin{align*}
&spec\ (monoTy\ T)\ nil\ T \triangleq \top \\
&(\nabla a. spec\ (polyTy\ \lambda a. P\, a)\ 
   (cons\ a\ (L\, a))\ (T\, a)) \triangleq \\
&\hspace{1cm}
      (\nabla a. \fresh a (L\, a)) \land
      (\nabla a. spec\ (P\, a)\ (L\, a)\ (T\, a))
\end{align*}
In this last clause we can again simplify the freshness condition
to produce the following.
\begin{align*}
&(\nabla a. spec\ (polyTy\ \lambda a. P\, a)\ (cons\ a\ L)\ (T\, a))
\triangleq \\
&\hspace{1cm} \nabla a. spec\ (P\, a)\ L\ (T\, a)
\end{align*}

\subsubsection{Capture-avoiding substitution}

The following program clauses realize capture avoiding substitution
for $\lambda$-terms via a predicate $subst(E, T, X, E')$ which holds
exactly when $E[T/X] = E'$.
\begin{align*}
&subst(var(X),E,X,E) \ent \top \\
&subst(var(x),E,y,var(x)) \ent \top \\
&subst(app(M,N),E,X,app(M',N')) \ent \\
&\hspace{1cm}
     subst(M,E,X,M') \land subst(N,E,X,N') \\
&subst(lam(\langle y\rangle R), E, X, lam(\langle y\rangle R'))
\ent \\
&\hspace{1cm} y \# X \land y \# E \land subst(R,E,X,R')
\end{align*}
These program clauses translate to the following definitional clauses:
\begin{align*}
&subst\ (var\ X)\ E\ X\ E \triangleq \top \\
&(\nabla x, y. subst\ (var\ y)\ (E\, x\, y)\ x\ (var\ y)) \triangleq \top \\
&subst\ (app\ M\ N)\ E\ X\ (app\ M'\ N') \triangleq \\
&\hspace{1cm} subst\ M\ E\ X\ M' \land subst\ N\ E\ X\ N' \\
&(\nabla y. subst\ (lam\ \lambda y. R\, y)\ (E\, y)\ (X\, y)\ (lam\
\lambda y. R'\, y)) \triangleq \\
&\hspace{1cm} (\nabla y. \fresh y (X\, y)) \land
              (\nabla y. \fresh y (E\, y)) \land~ \\
&\hspace{1cm} (\nabla y. subst\ (R\, y)\ (E\, y)\ (X\, y)\ (R'\, y))
\end{align*}
Simplifying the freshness constraints and removing vacuous
$\nabla$-binders in the last clause produces the following.
\begin{align*}
&subst\ (lam\ \lambda y. R\, y)\ E\ X\ (lam\ \lambda y. R'\, y)
  \triangleq \\
&\hspace{1cm} \nabla y. subst\ (R\, y)\ E\ X\ (R'\, y)
\end{align*}

\subsection{Correctness}

The soundness and completeness of our translation are shown by the
following results. We elide most details, but show the important
lemmas and interesting cases.

\begin{lemma}
\label{lem:fresh}
Let $a$ be a name and $t$ a ground $\alpha$Prolog term. Then $\models
a \# t$ holds if and only if $\lra \fresh {\phi(a)} {\phi(t)}$ has a
proof in \Gm.
\end{lemma}
\begin{proof}
Induction on $t$.
\end{proof}

\begin{lemma}
\label{lem:eq}
Let $t$ and $u$ be ground $\alpha$Prolog terms. Then $\models t
\approx u$ holds if and only if $\phi(t) = \phi(u)$.
\end{lemma}
\begin{proof}
Induction on $t$.
\end{proof}

We define the support of an $\alpha$Prolog term as the set of all
names which appear free relative to name-abstractions, \ie,
$\supp(t) = \{a \mid a \# t \mbox{ does not hold}\}$. This is
consistent with the definition of support for \Gm terms since
$\phi(\supp(t)) = \supp(\phi(t))$. For a substitution $\theta$ we
define $\phi(\theta) = \{ \phi(t)/\phi(x) \mid t/x \in \theta\}$
and $\supp(\theta) = \bigcup_{t/x \in \theta} \supp(t)$.

\begin{lemma}
\label{lem:term-subst}
Let $t$ be an $\alpha$Prolog term and $\theta$ a substitution,
then $\phi(t\theta) = \phi(t)\phi(\theta)$.
\end{lemma}
\begin{proof}
Induction on $t$.
\end{proof}

\begin{lemma}
\label{lem:goal-subst}
Let $G$ be an $\alpha$Prolog goal, $\theta$ a substitution, and
$\vec{a}$ a list of names such that $\supp(\theta) \cap
\{\vec{a}\} = \emptyset$, then $\phi_{\vec{a}}(G\theta) =
\phi_{\vec{a}}(G)\phi(\theta)$.
\end{lemma}
\begin{proof}
Induction on $G$. Consider when $G = b \# t$. Then
$\phi_{\vec{a}}(G\theta) = \nabla \vec{a} . \fresh {\phi(b\theta)}
{\phi(t\theta)} = \nabla \vec{a} . (\fresh {\phi(b)}
{\phi(t)})\phi(\theta)$. Since $\supp(\theta) \cap \{\vec{a}\} =
\emptyset$, we can move the substitution outside of the
$\nabla$-binder to obtain $(\nabla \vec{a} . (\fresh {\phi(b)}
{\phi(t)}))\phi(\theta) = \phi_{\vec{a}}(G)\phi(\theta)$.
\end{proof}

\begin{theorem}
Let $\Delta$ be a set of program clauses, $G$ a ground goal, and
$\vec{a}$ a list of distinct names. There is a proof of $\Delta \Lra
G$ if and only if there is a proof of $\lra \phi_{\vec{a}}(G)$
assuming the definitional clauses $\phi(\Delta)$ and the clause for
{\sl fresh}.
\end{theorem}
\begin{proof}
In the forwards direction, the proof is by induction on the height
of the $\alpha$Prolog proof. First consider when the proof ends
with {\sc fresh} so that $G = b \# t$. Then it must be that
$\models b \# t$ which means $\lra \fresh {\phi(b)} {\phi(t)}$ has
a proof. Thus $\lra \phi_{\vec{a}}(G)$ which is $\lra \nabla
\vec{a} . \fresh {\phi(b)} {\phi(t)}$ also has a proof. The cases
for {\sc true}, {\sc equal}, {\sc and}, and {\sc or} are
similarly easy.

Suppose the proof ends with {\sc exists} so that $G = \exists
X.G'$ and $\Delta \Lra G'[t/X]$ has a proof for some $t$. By
induction we know $\lra \phi_{\vec{a}}(G'[t/X])$ has a proof. We
would like to move the substitution outside of $\phi$, but we
cannot do so unless the support of the substitution is disjoint
from $\vec{a}$. Abusing notation, we do this by splitting the
substitution into two parts. That is, we know $\lra
\phi_{\vec{a}}(G'[X\, \vec{a}/X][\lambda \vec{a}.t/X])$ has a
proof. Now the $\vec{a}$ are not free in the second substitution
and thus we can apply Lemma~\ref{lem:goal-subst} to conclude that
$\lra \phi_{\vec{a}}(G'[X\, \vec{a}/X])[\lambda
\vec{a}.\phi(t)/\phi(X)]$ has a proof. Thus $\lra \exists
X.\phi_{\vec{a}}(G'[X\, \vec{a}/X])$ has a proof and this is the
same as $\lra \phi_{\vec{a}}(\exists X.G')$. The case for {\sc
backchain} is similar in spirit, but more complex in the details.

Lastly, suppose the proof ends with {\sc new} so that $G = \new
b.G'$ and $\Delta \Lra G'$ has a proof. By induction $\lra
\phi_{\vec{a}b}(G')$ also has a proof which is the same as $\lra
\phi_{\vec{a}}(\new b.G')$.

In the backwards direction, the proof is by induction on the
height of the \Gm proof with a nested induction on the size of
$G$. Consider first when $G$ is $t \approx u$. We assume $\lra
\phi_{\vec{a}}(t \approx u)$ has a proof which means that $\lra
\nabla \vec{a}. \phi(t) = \phi(u)$ has a proof and thus $\lra
\phi(t) = \phi(u)$ also does. Therefore $\models t \approx u$ is
true and $\Delta \Lra t \approx u$ has a proof. The cases for when
$G$ is $\top$, a freshness relation, a conjunction, or a
disjunction are similarly easy.

Suppose $G = \exists X.G'$ so that $\lra \phi_{\vec{a}}(G'[X\,
\vec{a}/X])[t/X]$ has a proof for some $t$. It must be that $t =
\lambda\vec{a}. \phi(u)$ for some $u$. Since the support of the
outer substitution is disjoint from $\vec{a}$ we can move it
inside to know $\lra \phi_{\vec{a}}(G'[X\,
\vec{a}/X][\lambda\vec{a}.u/X])$ has the same proof and this is
just $\lra \phi_{\vec{a}}(G'[u/X])$. By induction $\Delta \Lra
G'[u/X]$ has a proof and thus $\Delta \Lra \exists X.G'$ has a
proof. Again, the case for when $G$ is a predicate is similar in
spirit, but more complex in the details.

Finally suppose $G = \new b.G'$ so that $\lra \phi_{\vec{a}b}(G')$
has a proof. By the inner induction hypothesis, $\Delta \Lra G'$
has a proof and thus so does $\Delta \Lra \new b.G'$.
\end{proof}

\subsection{Extending the translation}

We now drop the name-restriction on $\alpha$Prolog and allow $a$ and
$b$ to be arbitrary terms in expressions of the form $a \# t$, $(a\
b)\cdot t$, and $\langle a \rangle t$. The translation is easily
extended to this richer language. Goals of the form $u \# t$ translate
to $\fresh u t$ as before. Terms which do not satisfy the
name-restriction are first simplified so that all non-name-restricted
swappings and name-abstractions appear at the top level of an equality
goal. Then these translate to distinguished predicates which
implement swapping and name-abstraction. This simplification is only
needed statically since instantiations during proof search can only
contain ground terms, thus ensuring that non-name-restricted terms do
not appear dynamically.

To simplify a non-name-restricted term of the form $(u_1\ u_2)\cdot t$ or
$\langle u\rangle t$ we replace it with a fresh variable, say
$t'$, conjoin the distinguished goal $t' \mathrel{\dot{\approx}}
(u_1\ u_2)\cdot t$ or $t' \mathrel{\dot{\approx}} \langle u\rangle
t$ respectively, and quantify $t'$ appropriately (existentially
when replacing terms in the body, universally when replacing terms
in the head of a program clause). This forces all terms
to be name-restricted except the top level of terms occurring on
the right side of the $\mathrel{\dot{\approx}}$ relation. We
extend the translation to deal with this relation as follows:
\begin{align*}
\phi_{\vec{a}}(t' \mathrel{\dot{\approx}} (u_1\ u_2)\cdot t)
  &= \nabla\vec{a}. \swap {u_1} {u_2} t t' \\
\phi_{\vec{a}}(t' \mathrel{\dot{\approx}} \langle u\rangle t)
  &= \nabla\vec{a}. \abst u t t'
\end{align*}
Where {\sl swap} and {\sl abst} are defined by the following:
\begin{align*}
&\forall E. [(\nabla x,y. \swap x y {(E\, x\, y)} {(E\, y\, x)})
    \triangleq \top] \\
&\forall E. [(\nabla x. \swap x x {(E\, x)} {(E\, x)}) \triangleq
    \top] \\
&\forall E. [(\nabla x. \abst x {(E\, x)} (\lambda x. E\, x)
    \triangleq \top]
\end{align*}

In practice it seems that non-name-restricted swappings and
name-abstractions are exceedingly rare. Still, it is reassuring
that such detailed manipulations of variables and binding
are so succinctly described in \Gm.

The following lemmas show that our manipulations of formulas are
sound and that {\sl swap} and {\sl abst} correctly capture swapping
and name-abstraction, and therefore the extended translation can be
shown to be sound and complete.

\begin{lemma}
If $t \approx u$ and $\Delta[t/x] \Lra G[t/x]$ has a proof then so does
$\Delta[u/x] \Lra G[u/x]$.
\end{lemma}
\begin{proof}
Induction on the height of the proof.
\end{proof}

\begin{lemma}
Let $t'$ and $(a\ b)\cdot t$ be ground nominal terms. Then $\models t'
\approx (a\ b)\cdot t$ holds if and only if $\lra \swap a b t t'$ is
provable in \Gm.
\end{lemma}
\begin{proof}
Induction on $t$.
\end{proof}

\begin{lemma}
Let $t'$ and $\langle a\rangle t$ be ground nominal terms. Then $\models t'
\approx \langle a\rangle t$ holds if and only if $\lra \abst a t t'$ is
provable in \Gm.
\end{lemma}
\begin{proof}
Induction on $t$.
\end{proof}

\section{Higher-order Judgments}
\label{sec:hoj}

Higher-order judgments are a common and powerful tool for
specifications using higher-order abstract syntax. Among other things,
they have nice properties which are often used when reasoning about
such specifications. These properties do not directly hold for \Gm
specifications, so one may question if we have to give them up in any
system which admits translations from nominal logic specifications.
This is not the case: higher-order judgments can be encoded in \Gm
while keeping their nice properties. Although this does not provide a
direct connection between nominal logic specifications and
higher-order judgments, it does show how the expressiveness and
benefits of both approaches can coexist.

Our translation produces specifications
using {\em weak higher-order abstract syntax}, \ie, using
abstractions only at distinguished variable types. With higher-order
judgments it is much more common to
use {\em full higher-order abstract syntax}, \ie, where
abstraction is used at the same type as the terms being
constructed. For example, $\lambda$-terms would be represented
using only the following two constants.
\begin{align*}
app &: tm \to tm \to tm &
lam &: (tm \to tm) \to tm
\end{align*}
This representation provides a free notion of capture-avoiding
substitution based on meta-level $\beta$-reduction whereas 
a weak higher-order abstract syntax encoding would provide only
variable for variable substitution.
The following example defines evaluation for $\lambda$-terms using the
free notion of capture-avoiding substitution.
\begin{align*}
&eval\ (lam\ \lambda x.R\, x)\ (lam\ \lambda x.R\, x)
  \triangleq \top \\
&eval\ (app\ M\ N)\ V \triangleq \\
&\hspace{1cm}
  \exists R. eval\ M\ (lam\ \lambda x.R\, x) \land
  eval\ (R\, N)\ V
\end{align*}
With full higher-order abstract syntax, we can still
distinguish variables from other terms by using a definition
such as
$(\nabla x. \name x) \triangleq \top$ which holds only on nominal
constants. For example, the specification of type checking for
$\lambda$-terms represented using full higher-order abstract syntax
can be written as follows:
\begin{align*}
&tc\ G\ X\ T \triangleq \name X \land lookup\ X\ T\ G \\
&tc\ G\ (app\ E_1\ E_2)\ T' \triangleq \\
&\hspace{1cm}
   \exists T. tc\ G\ E_1\ (arr\ T\ T') \land tc\ G\ E_2\ T \\
&tc\ G\ (lam\ \lambda x. E\, x)\ (arr\ T\ T') \triangleq \\
&\hspace{1cm} \nabla x. tc\ (bind\ x\ T\ G)\ (E\, x)\ T'
\end{align*}
Thus in \Gm we have the same naturalness and expressiveness
with full higher-order abstract syntax as with the weaker version.

Moving to higher-order judgments, the specification for type checking
$\lambda$-terms can be written as follows in $\lambda$Prolog
\cite{nadathur88iclp}:
\begin{align*}
&tc\ (app\ M\ N)\ B \ent
  (tc\ M\ (arr\ A\ B) \land tc\ N\ A) \\
&tc\ (lam\ \lambda x.R\, x)\ (arr\ A\ B) \ent
  (\forall x. tc\ x\ A \Rightarrow tc\ (R\, x)\ B)
\end{align*}
Here $tc$ does not carry around an explicit typing context. Instead,
the context in $\lambda$Prolog is used to remember typing assignments
for bound variables: the $\forall$-quantifier encodes
fresh variable names and the $\Rightarrow$ connective encodes
hypothetical assumptions.
Besides elegance, the real benefit of
this encoding is that the underlying logic enjoys meta-theoretic
properties such as cut-admissibility and the preservation of
provability under instantiations for universal variables. As a
result, if one wants to reason about this specification they
obtain a free object-level substitution result for $tc$ which says,
roughly, if $tc\ (lam\ \lambda x. R\, x)\ (arr\ A\ B)$ and $tc\ N\ A$
are both derivable, then $tc\ (R\, N)\ B$ is also derivable. Such
substitution lemmas are quite common and useful, for example, in
showing that evaluation preserves typing.

Higher-order judgments do not allow one to
directly distinguish between free variables, and this can result
in awkward specifications in some instances. For example, the
following $\alpha$Prolog clauses specify a notion of inequality
over $\lambda$-terms:
\begin{align*}
&aneq\ (var\ a)\ (var\ b) \ent \top \\
&aneq\ (app\ M_1\ N_1)\ (app\ M_2\ N_2) \ent aneq\ M_1\ M_2
  \\
&aneq\ (app\ M_1\ N_1)\ (app\ M_2\ N_2) \ent aneq\ N_1\ N_2
  \\
&aneq\ (lam\ \langle a\rangle R_1)\ (lam\ \langle a\rangle R_2)
  \ent aneq\ R_1\ R_2 \\
&aneq\ (var\ X)\ (app\ M\ N) \ent \top \\
&\vdots
\end{align*}
There is no equally natural way to express this with higher-order
judgments due to the need to distinguish between variables in the
first clause (see \cite{cheney08toplas} for an example encoding in
$\lambda$Prolog). In fact, the ability to distinguish between
variables is fundamentally at odds with the idea of a free
substitution property for variables since such substitutions may
cause two different variables to be instantiated to the same term.
Thus, higher-order judgments necessarily lack a degree of
naturalness for some specifications.

It is important to note that in a dependently-typed system like LF
higher-order abstract syntax and higher-order judgments collapse
into a single notion \cite{harper93jacm}. While this provides for
a very elegant system, we should not let it confuse us between
these two notions. As our translation has shown, higher-order
abstract syntax allows for specifications that are at least as
natural and expressive as what is possible in $\alpha$Prolog,
while higher-order judgments may occasionally fall short.

We now propose a specification methodology which allows one to use
the full natural expressiveness of definitions in \Gm while still
being able to take advantage of higher-order judgments so that we
can benefit from their elegance and associated properties when
reasoning. The idea is to encode an interpreter for higher-order
judgments as a definition in \Gm and use this to encode particular
higher-order judgment specifications. In the full logic \G which
is used for reasoning, one can prove general instantiation and
cut-admissibility properties for the encoding of higher-order
judgments. These properties are then inherited for free by any
specification written using higher-order judgments.

For simplicity of presentation we consider only the second-order
fragment of $\lambda$Prolog. It is possible to encode full
higher-order $\lambda$Prolog, but second-order is sufficient for
the majority of examples. The encoding of second-order
$\lambda$Prolog into \Gm is presented in Figure~\ref{fig:seq}. In
this encoding $::$ is an infix constructor for lists, and
$\langle\cdot\rangle$ is used to distinguish atomic formulas.
Since we are considering only second-order $\lambda$Prolog, we
assume $A$ is atomic in $A \Rightarrow B$. The formula $\seq L G$
will hold when the $\lambda$Prolog formula $G$ is provable from
the atomic assumptions in $L$ and the clauses of our particular
specification.
These latter
clauses are encoded via the predicate {\sl prog} which holds on
the head and body of each encoded clause. For example, the clauses
for the $tc$ predicate are encoded into the following {\sl prog}
clauses:
\begin{align*}
\hbox{\sl prog}~ &{(tc\ (app\ M\ N)\ B)} \\
      &{(\langle tc\ M\ (arr\ A\ B)\rangle \land
         \langle tc\ N\ A\rangle)} \triangleq \top \\[5pt]
\hbox{\sl prog}~ &{(tc\ (lam\ \lambda x.R\, x)\ (arr\ A\ B))} \\
      &{(\forall x. tc\ x\ A \Rightarrow
         \langle tc\ (R\, x)\ B\rangle)} \triangleq \top
\end{align*}

\begin{figure}[t]
\begin{align*}
&\member B (B::L) \triangleq \top \\
&\member B (C::L) \triangleq \member B L \\[5pt]
&\seq L \top \triangleq \top \\
&\seq L (B \land C) \triangleq \seq L B \land \seq L C \\
&\seq L (B \lor C) \triangleq \seq L B \lor \seq L C \\
&\seq L (A \Rightarrow B) \triangleq \seq {(A :: L)} B \\
&\seq L (\forall x.Bx) \triangleq \nabla x. \seq L (B\, x) \\
&\seq L \langle A \rangle \triangleq \member A L \\
&\seq L \langle A \rangle \triangleq
  \exists B. \prog A B \land \seq L B
\end{align*}
\caption{Second-order $\lambda$Prolog in \Gm}
\label{fig:seq}
\end{figure}

The {\sl seq} encoding of second-order $\lambda$Prolog retains the
desirable properties of the logic which we formally state below.
Moreover, these properties can be proven completely within the full
logic \G.

\begin{lemma}[Instantiation]
\label{lem:inst}
Let $c$ be a nominal constant and $t$ a term of the same type. If
$\lra \seq L G$ then $\lra \seq {L[t/c]} {G[t/c]}$.
\end{lemma}

\begin{lemma}[Cut admissibility]
\label{lem:cut}
If $\lra \seq {(A::L)} G$ and $\lra \seq L {\langle A\rangle}$
then $\lra \seq L G$.
\end{lemma}

\begin{lemma}[Monotonicity]
\label{lem:monotone}
If $\lra \seq L G$ and every element of $L$ appears in $K$ then
then $\lra \seq K G$.
\end{lemma}

When working with full higher-order abstract syntax,
Lemmas~\ref{lem:inst} and \ref{lem:cut} are quite powerful and provide
the object-level substitution lemmas described earlier. Although a
definition like {\sl seq} could be encoded in $\alpha$Prolog or
a similar language, the corresponding lemmas would not be as useful
because of the weaker notion of substitution.

A fundamental restriction of the proposed method for specification
is that while \Gm definitions can make use of higher-order judgments,
it is not possible for higher-order judgments to use \Gm
definitions. This restriction is inherent in our encoding via {\sl
seq} and is necessary to preserve the desirable meta-properties of
higher-order judgments. Note, however, that this is a restriction
on ``control-flow'' and not ``data-flow'' since we may still use
higher-order judgments to specify a value which is then fed into a
\Gm specification.

\section{Conclusions and Future Work}
\label{sec:conc}

Our translation provides a direct and concrete connection between
$\alpha$Prolog and \Gm including a tight mapping from the devices of
the former to the corresponding devices of the latter. In particular,
our translation provides an understanding of the relationship between
the $\new$ and $\nabla$-quantifiers. When used at distinguished
variable types, the $\nabla$-quantifier exactly captures the meaning
of the $\new$-quantifier, at least from the perspective of
specification. However, the $\nabla$-quantifier can also be used at
types which contain other constructors which is essential for
full higher-order abstract syntax and which is not possible with
current understandings of the $\new$-quantifier. In addition, using
raising and the {\sl fresh} predicate, the $\nabla$-quantifier can be
freely moved up and down in a formula whereas the $\new$-quantifier is
always given a large scope since nominal logic does not have raising
and therefore cannot push the $\new$-quantifier underneath other
quantifiers.

Through our translation we have also shown that higher-order
abstract syntax specifications can have at least the same
naturalness and expressiveness as nominal logic specifications.
The resulting specifications are based on weak higher-order
abstract syntax, but we have argued that the same qualities can be
found with full higher-order abstract syntax. Therefore, despite
being a very high-level approach to binding, higher-order abstract
syntax can still be used naturally in specifications which demand
a fine-grained control over variables.

We have acknowledged the occasional failings of higher-order
judgments to naturally capture some aspects of specifications
involving binding. We have proposed a method which allows one to
use higher-order judgments when relevant and a stronger
specification language when preferred. This method allows one to
benefit from the elegance of higher-order judgment during
specifications and from their associated meta-properties during
reasoning.

Thus we have presented the logic programming language \Gm which has
the power of higher-order abstract syntax, the fine-grained variable
control of nominal specifications, and the ability to capture the
desirable properties of higher-order judgments.

It seems possible to develop a fairly direct reverse translation
from \Gm to $\alpha$Prolog in the cases where only weak higher-order
abstract syntax is used. We have not pursued this line of work since
full higher-order abstract syntax is more common and is required
to reap the complete benefits of using higher-order judgments. Towards
this, Gabbay and Cheney have developed a translation from
$FO\lambda^\nabla$, a first-order logic with $\lambda$-terms, full
higher-order abstract syntax, and the $\nabla$-quantifier, to a
variant of nominal logic with $\lambda$-terms and the
$\new$-quantifier \cite{gabbay04lics, cheney05fossacs}. Their
translation makes a similar connection between the $\new$- and
$\nabla$-quantifiers as in this paper though in the opposite
direction. However, the presence of $\lambda$-terms in their nominal
logic is unorthodox and it would be interesting to see a similar
result for a more traditional nominal logic.

We have ignored issues of executing specifications in our
discussions, but we consider them briefly now. Most $\alpha$Prolog
specifications can be efficiently executed \cite{cheney08toplas,
urban04tcs}, while some require an expensive operation known as
{\em equivariant unification} to backchain on clauses
with $\new$-quantifiers in the head \cite{cheney09jar}. It
should also be possible to efficiently execute \Gm definitions in
a similar way, with similar issues when unfolding definitional
clauses with $\nabla$-quantifiers in the head. The difficulty of
this corresponds roughly with that of solving equivariant
unification problems in $\alpha$Prolog. However, specifications in
\Gm tend to use $\nabla$-quantifiers in the head of definitions
less often than $\alpha$Prolog uses $\new$-quantifiers in the head
of clauses since \Gm has real $\lambda$-binders whereas
$\alpha$Prolog must use $\new$-quantification and name-abstraction
to represent a binder. To efficiently execute such benign uses of
the $\new$-quantifier in the head of clauses, researchers
have studied the notion of {\em $\new$-goal formulas}
\cite{cheney08toplas, urban05tlca} which are essentially those
which translate to \Gm definitions without $\nabla$-quantifiers in
the head of clauses except for the distinguished {\sl fresh}, {\sl
swap}, and {\sl abst} predicates.

In the worst case, our translation may produce a quadratic increase in
the size of formulas due to raising. In practice, this does not appear
be an issue for several reasons. First, most specifications mention
few object variables per formula and thus the amount of raising
required is fairly limited. Second, by making dependencies explicit
via raising we are able to statically solve freshness constraints as
shown in the examples. This removes some raising and decreases the
number of atomic formulas in a definition. Third, we are not proposing
that specifications in \Gm be written as if translated from
$\alpha$Prolog. Specifications that instead use full higher-order
abstract syntax and the corresponding notion of substitution inherit
free implementation benefits. For example, an implementation can
lazily apply substitutions which may result in significant performance
improvements. Further research is needed to accurately assess the
relative efficiency of nominal and higher-order abstract syntax
specifications.

The Abella system \cite{gacek08ijcar} is a theorem prover for the
logic \G and supports the proposed method of specification where
higher-order judgments are mixed with \G definitions. In practice,
we have found that this hybrid style provides a nice compromise
between elegance and practicality. The reasoning over such
specifications benefits from the meta-properties of higher-order
judgments as expected and from the naturalness of specifications
which directly use the features of \G. Abella can also
execute these specifications, though it is not optimized for this.
Through the translation given in
this paper, it is possible to use Abella to execute and reason
about $\alpha$Prolog specifications. An interesting direction for
future work would be to assess such capabilities and to possibly
develop them explicitly within Abella.

Finally, we note that higher-order judgments are incorporated in
\G via a definition and not by using the universal quantifier and
implication of \G. Indeed, the latter devices have a much
different behavior in \G than in higher-order judgments. In
higher-order judgments, universal quantification denotes a generic
quantification and implication denotes a fixed assumption. In \G,
universal quantification denotes a quantification over each and
every possible value and implication restricts attention to worlds
in which the hypotheses are provable. The Bedwyr system is based on a
a \G-like logic and uses these devices to encode and execute
specifications of model-checking behavior such as bisimulation for the
finite $\pi$-calculus \cite{baelde07cade}.

\bibliographystyle{abbrvnat}
\bibliography{../references/master}

\end{document}